\newcommand{\bksg} {\mbox{$B^0 \! \rightarrow K^{*0}\gamma$}}
\newcommand{\brhog} {\mbox{$B^0 \! \rightarrow \rho^{0}\gamma$}}
\newcommand{\bpppkkk} 
{\mbox{$B^0\!\rightarrow \pi^+\pi^-,K^+\pi^-,K^+K^-$}}
\newcommand{\bpp} 
{\mbox{$B^0\!\rightarrow \pi^+\pi^-$}}
\newcommand{\bkp} 
{\mbox{$B^0\!\rightarrow K^+\pi^-$}}
\newcommand{\bkk} 
{\mbox{$B^0\!\rightarrow K^+K^-$}}
\newcommand{\bhh} 
{\mbox{$B^0\!\rightarrow h^+h^{'-}$}}

\documentclass[11pt]{article}
\usepackage{graphicx}
\usepackage{epsf}

\newcommand{\BABARPubYear}    {01}

\newcommand{\BABARProcNumber} {14}
\newcommand{\SLACPubNumber} {8848}
\newcommand{\LANLNumber} {0106043}

\input pubboard/babarsym

\long\def\inst#1{\par\nobreak\kern 4pt\nobreak
    {\it #1}\par\vskip 10pt plus 3pt minus 3pt}

\begin{document}
{\pagestyle{empty}

\begin{flushright}
SLAC-PUB-\SLACPubNumber \\
\babar-PROC-\BABARPubYear/\BABARProcNumber \\
hep-ex/\LANLNumber \\
June, 2001 \\
\end{flushright}

\par\vskip 4cm

\begin{center}
\Large \bf 
Charmless B Decays from \Lbabar\\\
(\bksg\ and \bpppkkk)
\end{center}
\bigskip

\begin{center}
\large 
M. Convery\\
Stanford Linear Accelerator Center\\
Stanford University, Stanford, CA 94309\\
(for the \lbabar\ Collaboration)
\end{center}
\bigskip \bigskip

\begin{center}
\large \bf Abstract
\end{center}
The BaBar experiment has completed its first year of data-taking 
during which 21 $fb^{-1}$ of data were accumulated. 
We present preliminary results on two types of charmless B decays
based on this data sample.
The first, \bksg, is  a so-called electromagnetic penguin.
And the second, \bhh, where $h^{(')}$ 
can be either a pion or kaon, can come from
either penguin diagrams or Cabibbo suppressed tree diagrams.

\vfill
\begin{center}
Contributed to the Proceedings of the 
XV Rencontres de Physique de La Vall\'ee d'Aoste, \\
March 4--10, 2001, La Thuile, Italy.
\end{center}

\vspace{1.0cm}
\begin{center}
{\em Stanford Linear Accelerator Center, Stanford University, 
Stanford, CA 94309} \\ \vspace{0.1cm}\hrule\vspace{0.1cm}
Work supported in part by Department of Energy contract DE-AC03-76SF00515.
\end{center}


\section{Theory and Motivation}
In the Standard Model, the charmless $B$ decays that will be discussed in
this paper proceed via the Feynman diagrams shown in Figure \ref{fig:feyn}.
The diagram shown in Figure \ref{fig:feyn}(a) is the ``penguin'' diagram.
This diagram has a loop with a $u$-type quark, which can emit either 
a $Z^0$, a photon ($\gamma$), or a gluon (g) before recombining to form
either $d$ or $s$ quark.
Due to its heavy mass, the diagram containing a top quark is much 
larger than those containing $u$ or $c$ quarks, which can thus
be ignored.
Therefore, there are CKM factors of $V_{ts}$ or $V_{td}$ at the
recombination vertex, 
depending on whether the final state contains an $s$ or $d$ quark.
If the boson emitted in the quark-loop is a photon, then final states
such as \bksg\ 
\footnote{Charge conjugation is assumed throughout this paper unless
specifically stated.}
(with an $s$-quark in the final state) and 
\brhog\ (with a $d$-quark in the final state) can be produced.
If the emitted boson is a gluon, then fully hadronic states 
such as \bkp\ can be produced.
Neither diagrams containing $Z^0$ bosons, nor those with $d$-quarks have
yet been observed.

Measuring decays produced by the penguin diagram is interesting because
it can potentially give a low energy window on high energy phenomena.
For example, in the Standard Model, measuring the branching ratio of 
\brhog\ will give information about the value of $V_{td}$.
Beyond the standard model, a charged Higgs ($H^+$) or similar particle 
could replace the $W$ and either enhance the branching ratio for 
\bksg\ or lead to direct CP violation \cite{joann}.

The ``tree diagram'' of Figure \ref{fig:feyn}(b) can lead to the 
decay \bpp.
This is especially interesting, because the final state is a CP-eigenstate
and the decay has a Matrix Element that is proportional to the the CKM 
factor $V_{ub}$.
This means that the decay could be used to measure time-dependent CP 
violation.
Such a measurement, when combined with constraints from isospin
conservation, could  yield
a very clean measurement of the unitarity angle 
$\alpha$, which is difficult to measure in any other way
\cite{gronau}.
However, the penguin diagram can also produce the same final state
and, since it has a different matrix element, could 
complicate the measurement.
Regardless, a precise measurement of $\alpha$ will require higher
statistics than are currently available.

In this paper, we will present results on analyses of the decay modes
\bksg\ and decays to two light charged hadrons (\bpppkkk), which we
will often call \bhh.
Preliminary BABAR results on charmless decays
based on about half of the current data set
have been presented in \cite{osakahh} (\bhh) and in 
\cite{osakaksg} (\bksg).
Here, we present updated preliminary results based on the full 
data-set of the first year.

\begin{figure}[t]
\begin{center}
\hbox{
\hbox{\epsfysize 2.3 truein \epsfbox{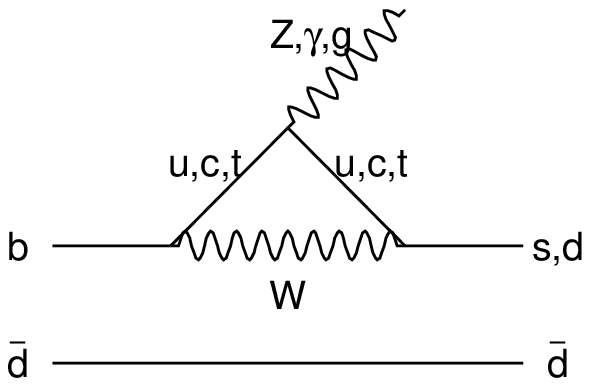}}
\kern-1.8in\lower 0.1in \hbox{(a)}
\kern1.8in
\hbox{\epsfysize 1.7 truein \epsfbox{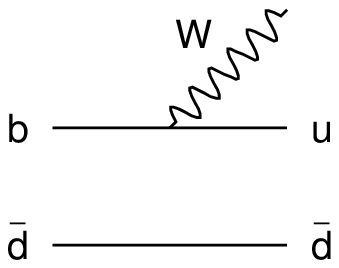}}
\kern-1.2in\lower 0.1in \hbox{(b)}
}
\end{center}
\caption{\it \label{fig:feyn} Feynman diagrams responsible for 
charmless $B$ decays. (a) is the ``penguin'' diagram, which
contributes to the all of the final states being discussed 
in this paper.
(b) is the ``tree'' diagram, which can contribute to the 
\bpp decay.
}
\end{figure}

\section{The PEP-II B Factory}
The data that will be described in this paper was taken with the BABAR
detector at the PEP-II B Factory, which is at located the Stanford 
Linear Accelerator Center (SLAC) and has been running since May of 1999.
PEP-II is an asymmetric $e^+e^-$ collider that collides a beam of $e^-$'s
with an energy 9.0 GeV with a beam of $e^+$'s with an energy of 3.1 GeV.
This produces a center of mass energy of equal to the 
$\Upsilon{\rm (4s)}$ mass and thus provides a copious source of $B$ mesons.
The peak luminosity achieved in ``Run 1'', which stretched from November
of 1999 to October of 2000, was $3.3 \times 10^{33} {\rm cm^{-2}sec{-1}}$.
The integrated luminosity collected during that time was 
approximately 21 $fb^{-1}$, which 
is equivalent to $(22.7 \pm 0.4)\times 10^6 \; B\bar{B}$ pairs.
Additionally, 3 $fb^{-1}$ of data was taken at energies below the 
$\Upsilon{\rm(4s)}$ peak.

The asymmetric energies of the PEP-II beams are required for doing the
time-dependent CP asymmetries that are the heart of the BABAR physics
program \cite{dave}.
However, they cause some problems for the measurement of charmless
branching ratios.
For example, the range of momenta of the daughters produced in two-body
B decays is rather narrow in the center of mass frame
\hbox{($2.4 \, {\rm GeV} < p_{2-body, CMS} < 2.8 \, {\rm GeV}$)}.
In the lab frame, however, the range is wider
($2 \, {\rm GeV} < p_{2body, lab} < 4 \, {\rm GeV}$).
This means, for example, that the the particle identification 
device used to separate 
$\pi$'s from $K$'s needs to work over a wider range of momenta
than it would otherwise.

\section{The BABAR Detector}
Figure \ref{fig:babar} shows a schematic of the BABAR detector.
The innermost component of BABAR is 5 layer Silicon Vertex Tracker.
Outside of that is a 40 layer central Drift Chamber.
The DIRC (Detector of Internally Reflected Cherenkov light)
\cite{dirc},
which
is used for charged hadron identification is composed of two parts:
the quartz radiator bars, which are located just outside of the 
Drift Chamber,
and the array of Photo Multiplier tubes,
which is located on the backward side
of detector outside the central region.
Surrounding the DIRC radiator bars is an electromagnetic calorimeter 
using CsI crystals.
Outside of this is the Superconducting Solenoid, which provides a 1.5 T
field.
Finally, the solenoid is surrounded by an iron flux return, which has 
gaps instrumented with resistive plate chambers to provide muon and neutral
hadron identification.
The capabilities of BABAR are described in more detail in \cite{dave}. 
Since the charged hadron identification provided by the DIRC is key to 
both of the analyses described in this note, we will describe it in more
detail here.

\begin{figure}
\begin{center}
\epsfysize 4.0 truein \epsfbox{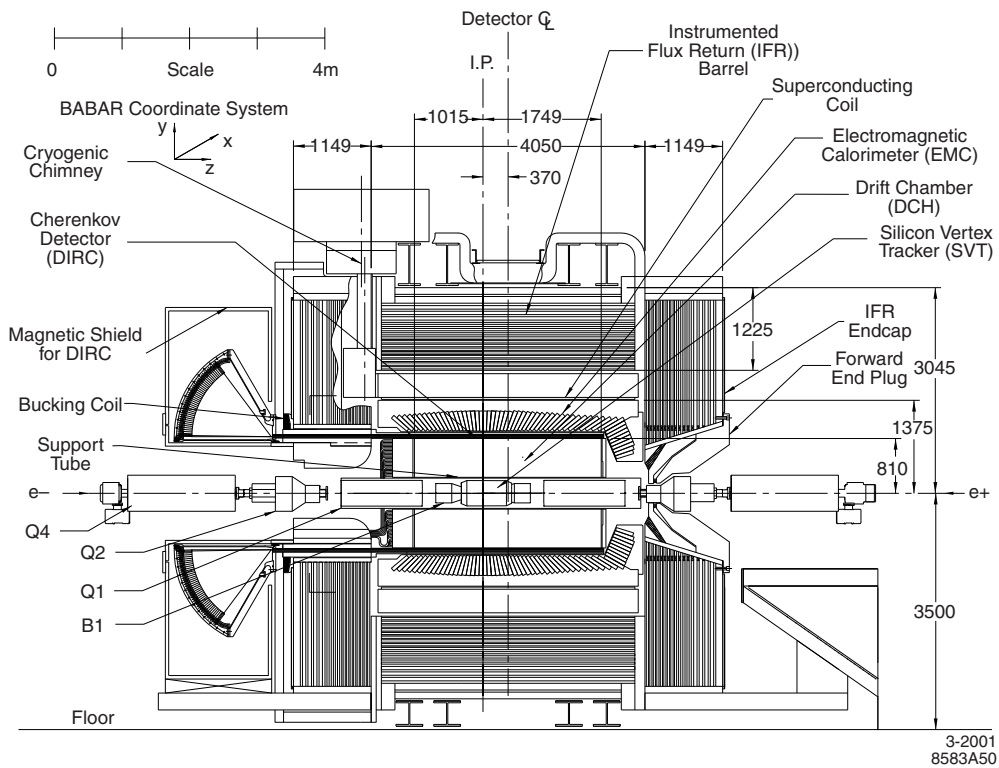}
\end{center}
\caption{\it \label{fig:babar}Side view of the BABAR detector.}
\end{figure}

\section{The DIRC}
Figure \ref{fig:dircprin} shows the principle of DIRC operation.
As charged particles pass through the radiator bars, Cherenkov light 
is produced. 
A fraction of that light is trapped inside the rectangular bars by 
total internal reflection.
As the light than propagates down the length of the bar, typically
bouncing 200 times, the Cherenkov pattern is preserved, up to reflection
ambiguities.
At the end of the bar, the light emerges into a stand-off region where
the Cherenkov pattern is allowed to expand before being detected by
an array of roughly 11,000 phototubes.

\begin{figure}
\begin{center}
\epsfysize 3.0 truein \epsfbox{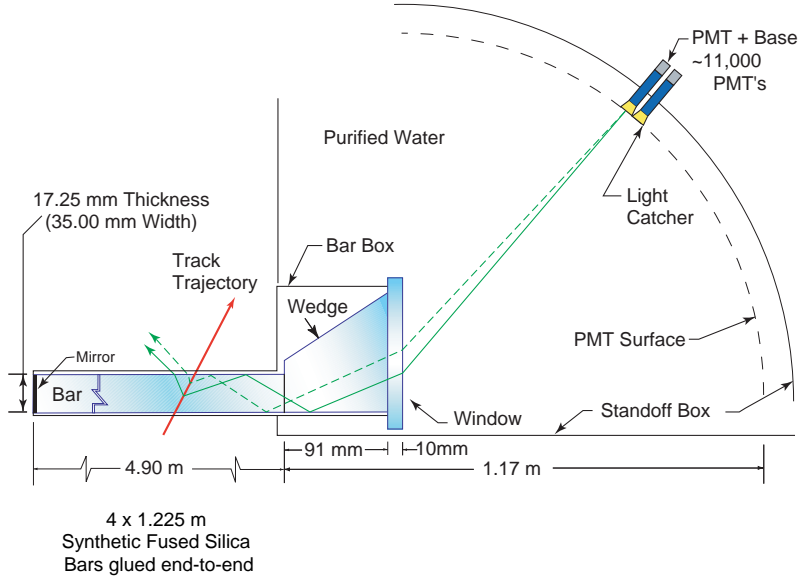}
\end{center}
\caption{\it \label{fig:dircprin}
The principle of DIRC operation. See text for explanation.
}
\end{figure}

Measurement of the Cherenkov angle of the detected light provides
excellent $\pi$, $K$ separation.
The performance of the DIRC can be checked using samples of 
$\pi$'s and $K$'s identified in 
$D^{*+}\!\rightarrow  \pi^+ D^0 (D^0\!\rightarrow K^-\pi^+)$.
Figure \ref{fig:pksep}(a) shows the performance of a $K$ selection
algorithm based on the DIRC Cherenkov angle measurement on these
samples.
The efficiency for correctly identifying $K$'s is roughly 90\%,
while the probability of mis-identifying a $\pi$ as $K$ is roughly 
2 \% at low momentum, but degrades somewhat at high ($>2.5 {\rm GeV}$)
momentum.
Figure \ref{fig:pksep}(b) shows the bands of Cherenkov angles measured
by the DIRC for high momentum tracks.
The bands grow closer together at high momentum, but the separation is 
still better than 2.5 $\sigma$ at 4 GeV, which is the kinematic limit
for two-body $B$ decays.

\begin{figure}
\begin{center}
\hbox{
\hbox{\epsfysize 2.3 truein \epsfbox{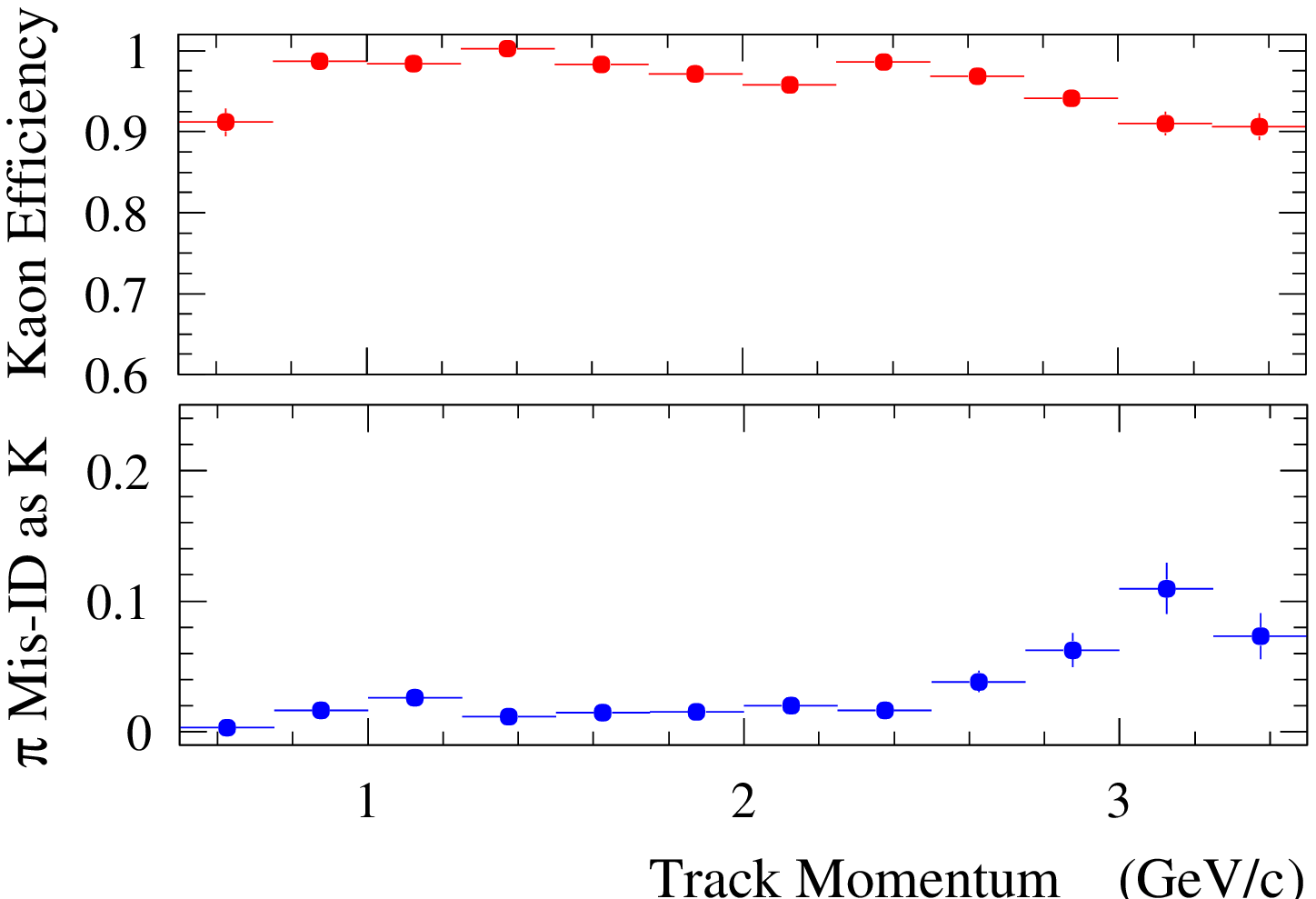}}
\kern-1.8in\lower 0.2in \hbox{(a)}
\kern1.8in
\hbox{\epsfysize 2.3 truein \epsfbox{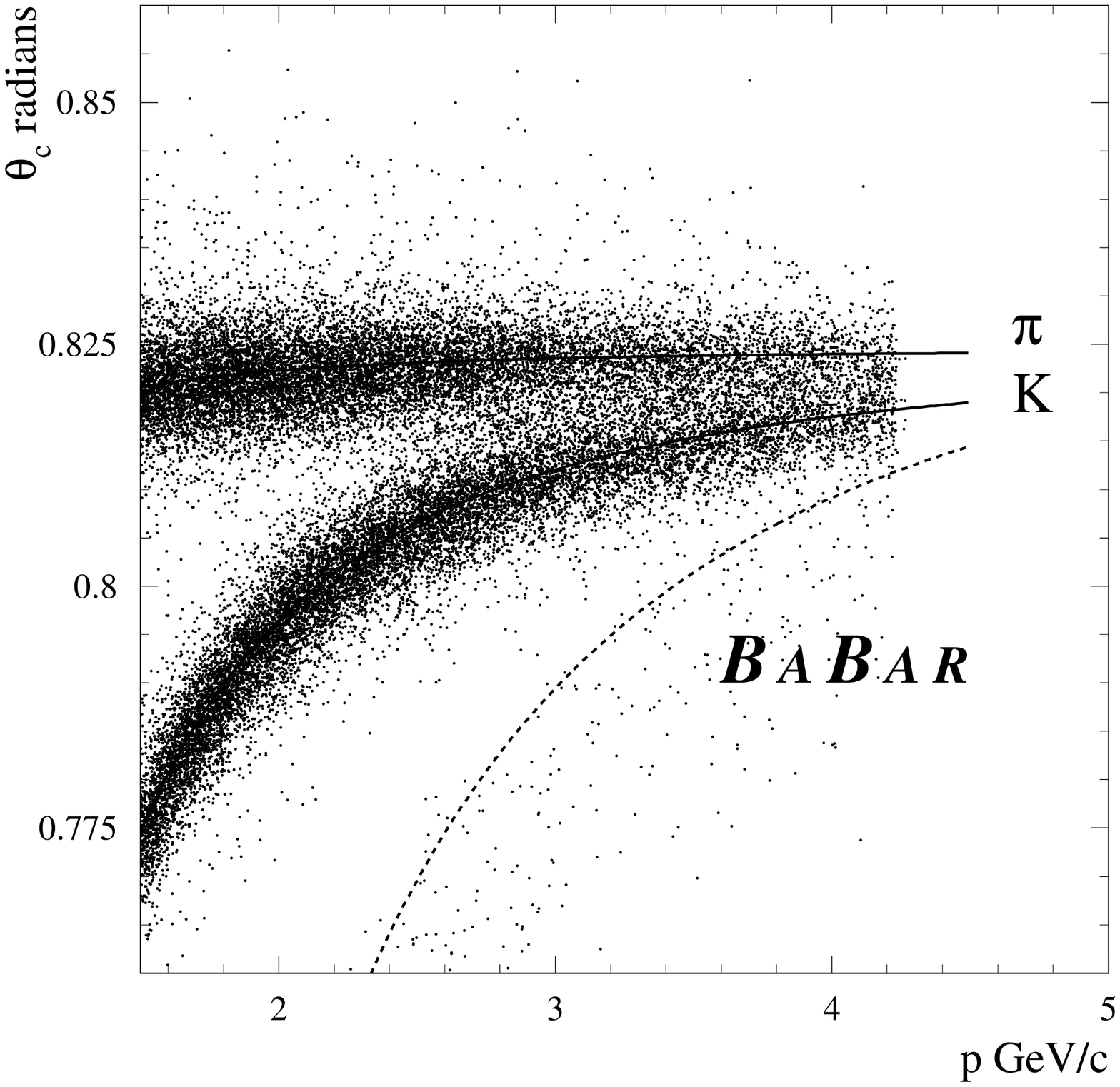}}
\kern-1.2in\lower 0.2in \hbox{(b)}
}
\end{center}
\caption{\it \label{fig:pksep}
$\pi$/K separation of the DIRC. a) shows the performance of a 
kaon selection algorithm using DIRC information.
b) shows the raw Cherenkov angle measurement for samples
of $\pi$'s and K's.
}
\end{figure}

\section{Analysis Procedure}
\label{sec:analproc}
A very similar procedure is used for both the \bksg\ and \bhh\
analyses.
The first step is to compose ``B candidates'' of the desired
final state out of the available charged tracks, photons and 
$\pi^0$'s.
Then, background is rejected based on event shape variables.
Background levels are measured using off-resonance data or
sideband regions of on-resonance data.
To determine the number of signal events, a fit is performed to 
a set of kinematic variables.
The signal efficiency is calculated in Monte Carlo, including 
adjustments for tracking efficiency and other related effects.

\section{Background Suppression}
\label{sec:bkgsupp}
Because the cross-section for continuum light-quark production
is much higher than that for $B\bar{B}$ production
$\sigma(e^+e^-\! \rightarrow q\bar{q}) \approx 
3\sigma(e^+e^-\! \rightarrow B\bar{B})$
and because the tracks coming from continuum events tend to have 
higher momentum, the background to charmless decays comes mostly from
continuum events.
This background may be suppressed by noting that
in the center of mass frame, continuum events are more ``jetty'' than 
the more spherical $B\bar{B}$ events.

The first step in this procedure is to define the ``axis'' of the candidate
B-decay. 
In the case of \bksg, this is  taken to be the direction of the photon.
While for the case of \bhh, it is taken as the thrust axis of the
B candidate.
Then, the tracks and neutral energy of the the remainder of the event 
(excluding the B candidate) are examined.
In the case of \bksg\, the thrust axis of the remainder is calculated
and the angle between it and the candidate axis is called $\theta_t$.
Figure \ref{fig:bkg}(a) shows the distribution of the 
$\cos \theta_t$ for signal and background.
As expected for the ``jetty'' background, $\cos \theta_t$ is peaked toward
1.
For the more spherical $B\bar{B}$ events, it has a flat distribution.

In the case of \bhh\, a more sophisticated approach, which was originally
developed by CLEO is used \cite{cleo}.
This method defines a set of 9 concentric cones centered on the candidate 
axis.
Each cone subtends $10^\circ$ of solid angle and is folded to combine
the forward and backward intervals.
The energy in the $i$th cone is called $x_i$ and a ``Fisher Discriminant'',
$F$,
is defined as
\begin{equation}
F = \sum_{i=1}^9 \alpha_i x_i.
\end{equation}
The coefficients $\alpha_i$ are varied so as to maximize the separation
between signal and background.
Figure \ref{fig:bkg}(b) shows histograms of this quantity for signal
and background.

\begin{figure}
\begin{center}
\hbox{
\hbox{\epsfysize 2. truein \epsfbox{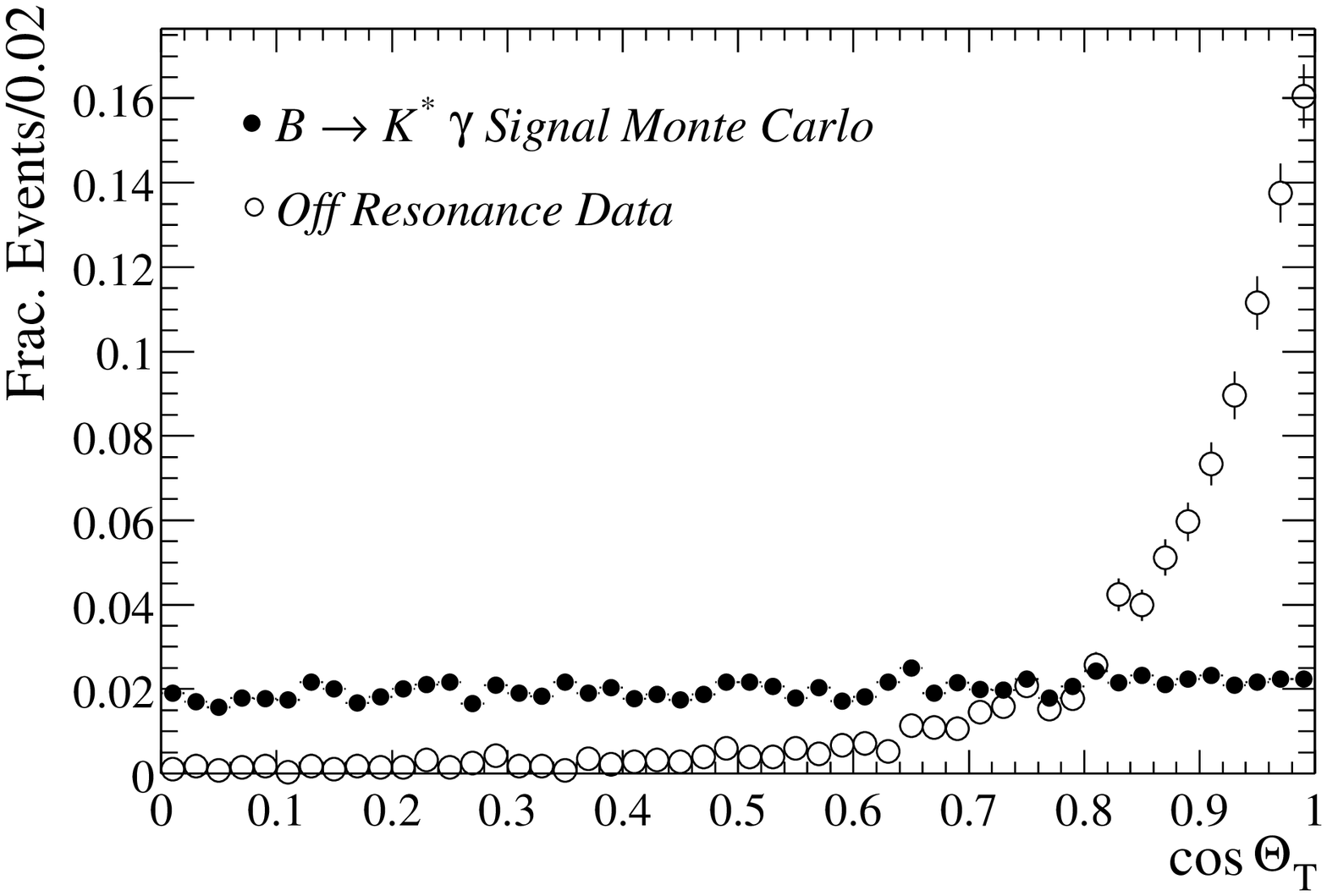}}
\kern-1.6in\lower 0.2in \hbox{(a)}
\kern1.8in
\lower 0.2in
\hbox{\epsfysize 2.3 truein \epsfbox{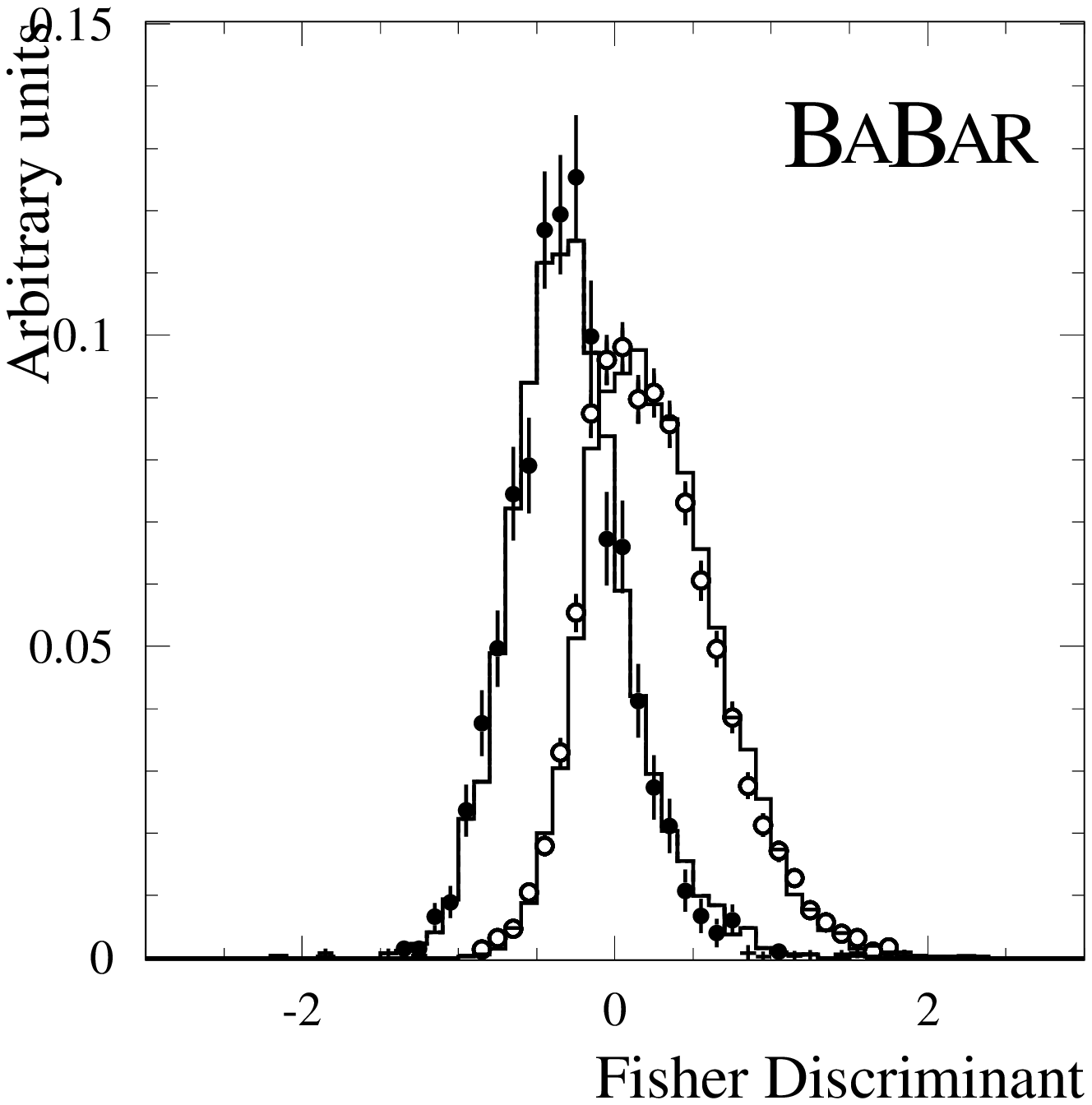}}
\kern-1.1in\lower 0.2in \hbox{(b)}
}
\end{center}
\caption{\it \label{fig:bkg}
Event shape variables used for background suppression.
a) shows the $\cos \theta_t$ variable used in \bksg.
b) shows the Fisher Discriminant used in \bhh. Solid circles are for
signal, open circles for background.
}
\end{figure}

\section{Kinematic Variables}
\label{sec:kinvar}
Two kinematic variables are traditionally used in 
exclusive $B$-reconstruction
in symmetric colliders operating at the $\Upsilon$(4s).
The first, known as the ``beam-constrained mass'' is defined as,
\begin{equation}
m_B \equiv \sqrt{E^2_{beam} - p_B^2},
\end{equation}
where $E_{beam}$ is the beam energy and $p_B$ is the momentum of the 
$B$ candidate in the center of mass system.
Since the beam energy is precisely known, this variable has much
better mass resolution than would be achieved by using the measured
energy of the $B$ candidate.
The second variable is defined  as
\begin{equation}
\Delta E \equiv E_B - E_{beam},
\end{equation}
where $E_B$ is the measured energy of the B candidate.
In the case of \bhh, if one assumes that both daughters have
the pion mass,
then $\Delta E$ will peak at zero for signal events for which
both daughters are pions.
If one or more of the daughters are not pions, 
then the peak will be shifted away 
from zero, thus providing particle identification.
In the case of \bksg, where there is no ambiguity in the particle masses,
one expects a peak at zero.

In an asymmetric collider, one can simply boost the track parameters back
into the center of mass frame and calculate $m_B$ as usual.
However, this requires assignment of masses to each of the particles.
In some analyses, such as \bhh, this is undesirable and can 
lead to worse $m_B$ resolution if the wrong mass assignment is made.
In such cases, it is desirable to define a new variable called
the energy-substituted mass  $m_{ES}$,
\begin{equation}
m_{ES} \equiv \sqrt{({1 \over 2}s + \vec{p_0} \cdot \vec{p_B})/E^2_0 -p_B^2},
\end{equation}
where $\sqrt{s}$ is the center of mass energy, $\vec{p_0}$ and $E_0$ 
are the three-momentum and energy of the $\Upsilon \rm{(4s)}$ and
$\vec{p_B}$ is the three-momentum of the $B$ candidate.
This variable has the advantage that it uses only quantities measured
in the lab system, and thus requires no mass assignments to be made.
It is identical to $m_B$ if the lab and center of mass frames coincide.
It is also equivalent to $m_B$ if there is no ambiguity in the 
particle masses.
A new version of $\Delta E$ is also defined,

\begin{equation}
\label{eq:delta}
\Delta E^* \equiv E_B^* - \sqrt{s}/2,
\label{eqn:deltae}
\end{equation} 
which also has shifts for different daughter masses.

\section{\bksg\ Analysis}
\subsection{\bksg\ Candidates}
The \bksg\ decay is searched for in the mode 
$K^{*0}\!\rightarrow K^{\pm}\pi^{\mp}$.
\bksg\ decay candidates are composed out of available photons and
$K^{*0}$'s, which are in turn composed of $K^{\pm}$'s and $\pi^{\mp}$'s.

In order to be considered, the photon candidate must be located
cleanly within the active region of the calorimeter 
($-0.73 < \cos\theta_{lab}<0.9$) and must have an energy close to 
that expected in the center of mass 
($2.3 \, {\rm GeV} < E_{\gamma,CMS} < 2.8 \, {\rm GeV}$).
In addition, the photon must not be consistent with having come
from a $\pi^0$ or $\eta$ decay.
Two cuts are applied to insure this.
The transverse shape of the shower is used to reject those
$\pi^0$'s for which the two photons are not spatially separated.
And, to reject $\pi^0$'s and $\eta$'s that
do form two separate calorimeter clusters,
the candidate photon is matched up with all other photons in the event
and rejected if any of them gives a mass consistent with 
$m_{\pi^0}$ or $m_\eta$.
The efficiency of these photons cuts is 77\%.

Similarly, the $K^{*0}$ candidate must pass a number of cuts.
First, the $K^+$ and $\pi^-$ candidates must pass DIRC particle 
identification cuts.
Then, the mass of the $K\pi$ system must be consistent
with the $K^{*0}$ ($0.796 {\rm GeV} < m_{K\pi} < 0.996 {\rm GeV}$).
Finally, the helicity angle, 
defined as the angle between the $K^+$ direction
in the $K^*$ rest frame and the $K^*$ flight direction, must satisfy
$|\cos \theta_{hel}|< 0.75$.
The efficiency of the $K^{*0}$ cuts is 56 \%.
Figure \ref{fig:kstarmass} shows the $K^{*0}$ mass peak in the final 
\bksg\ sample.

\begin{figure}
\begin{center}
\epsfysize 2.3 truein \epsfbox{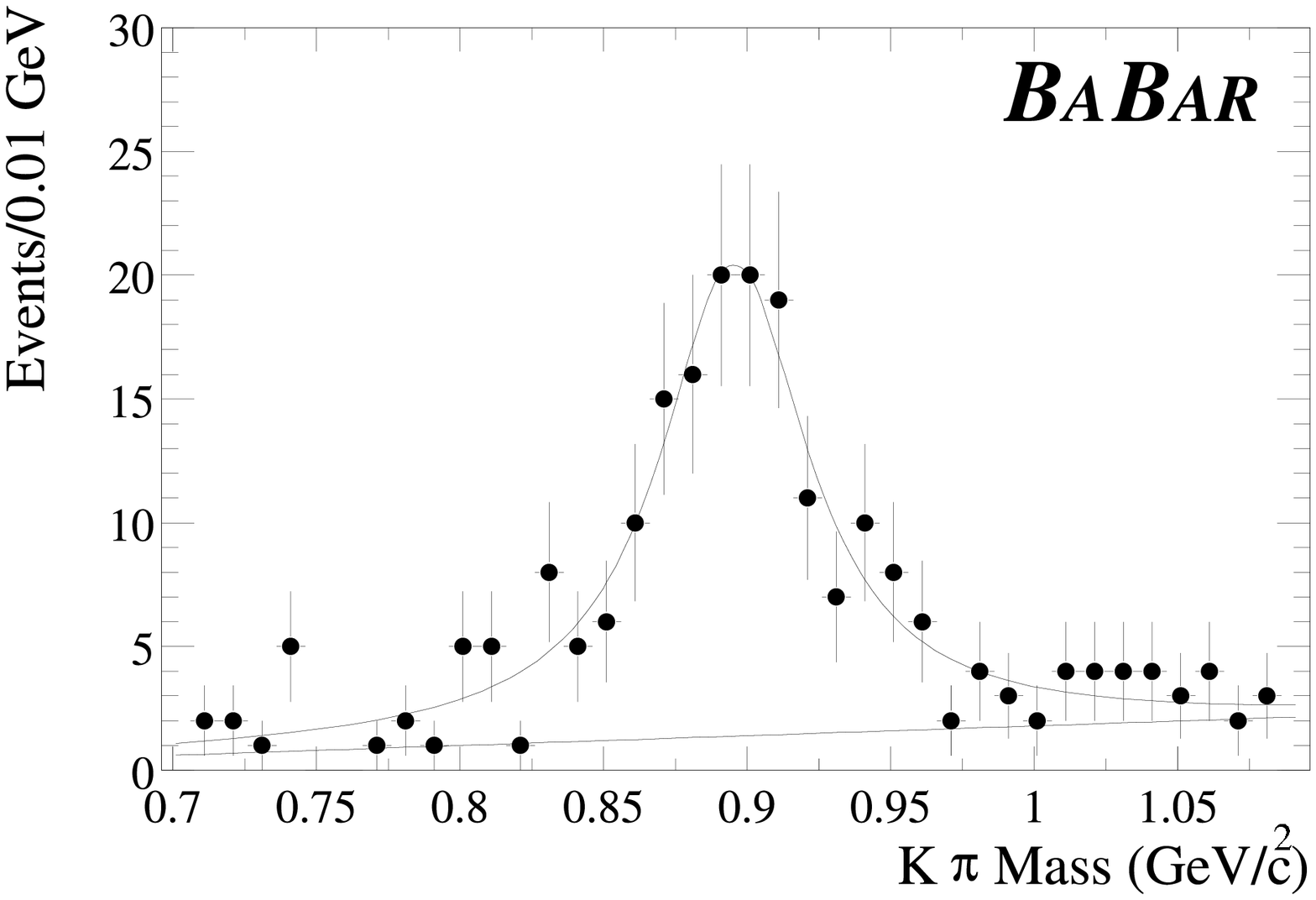}
\end{center}
\caption{\it \label{fig:kstarmass}
$K^{*0}$ mass peak for the final \bksg\ sample.
}
\end{figure}

\subsection{\bksg\ Mass Fit}
Once the photon and $K^{*0}$ candidates have been identified, $B$ candidates
are formed from them.
These candidates must pass a number of cuts in order to be considered in the
final sample.
The $\Delta E^*$ of the candidate, defined in Equation \ref{eq:delta} 
must pass a cut of $ -200 MeV <  \Delta E^* < 100 MeV $. 
The ``thrust angle'' must pass the cut 
$|\cos \theta_t| < 0.8$.
And finally, the polar angle of the B candidate in the center of mass frame
must pass the cut
$|\cos \theta_B^{*}| <$ 0.75.
The events that pass these cuts are shown in Figure \ref{fig:ksgmes}(a) and
(b). 
The concentration of events near $\Delta E^* = 0$ and 
$M_{B} = 5.28$ in Figure \ref{fig:ksgmes}(a)
indicates a strong \bksg\ signal.
In order to determine the number of signal events, a fit is performed on the
$M_{B}$ distribution. 
As shown in Figure \ref{fig:ksgmes}(b), he background is modeled with 
an ``Argus'' function, whose shape is determined using data taken off the 
$\Upsilon (4s)$ resonance. 
The shape of the signal is Gaussian, and the mean, $\sigma$ and amplitude
are floated in the fit.
The number of signal \bksg\ events extracted from the fit is 
$N_{sig} = 139.2 \pm 13.1$.

\begin{figure}
\begin{center}
\hbox{
\hbox{\epsfysize 1.9 truein \epsfbox{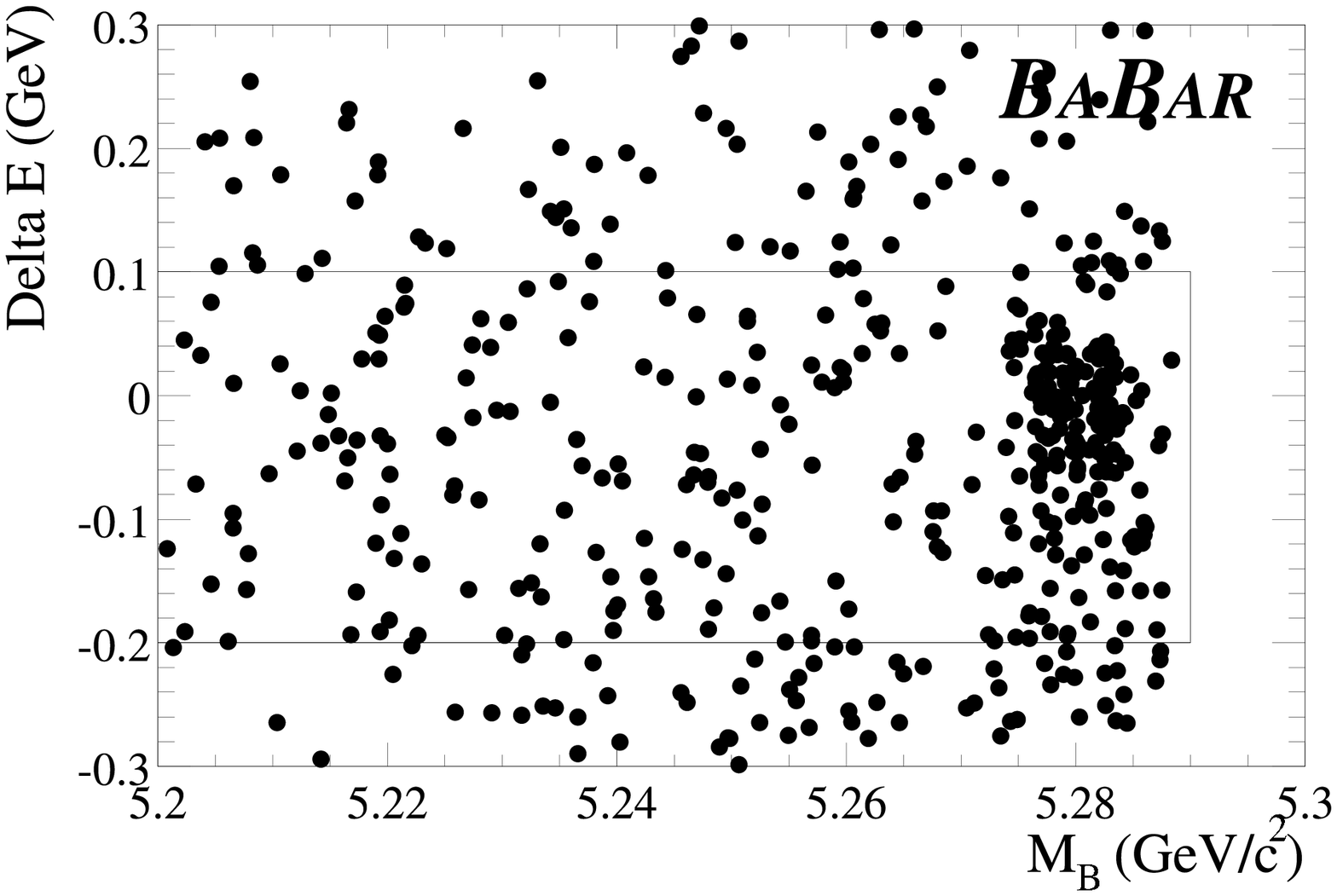}}
\kern-1.5in\lower 0.1in \hbox{(a)}
\kern1.3in
\hbox{\epsfysize 1.9 truein \epsfbox{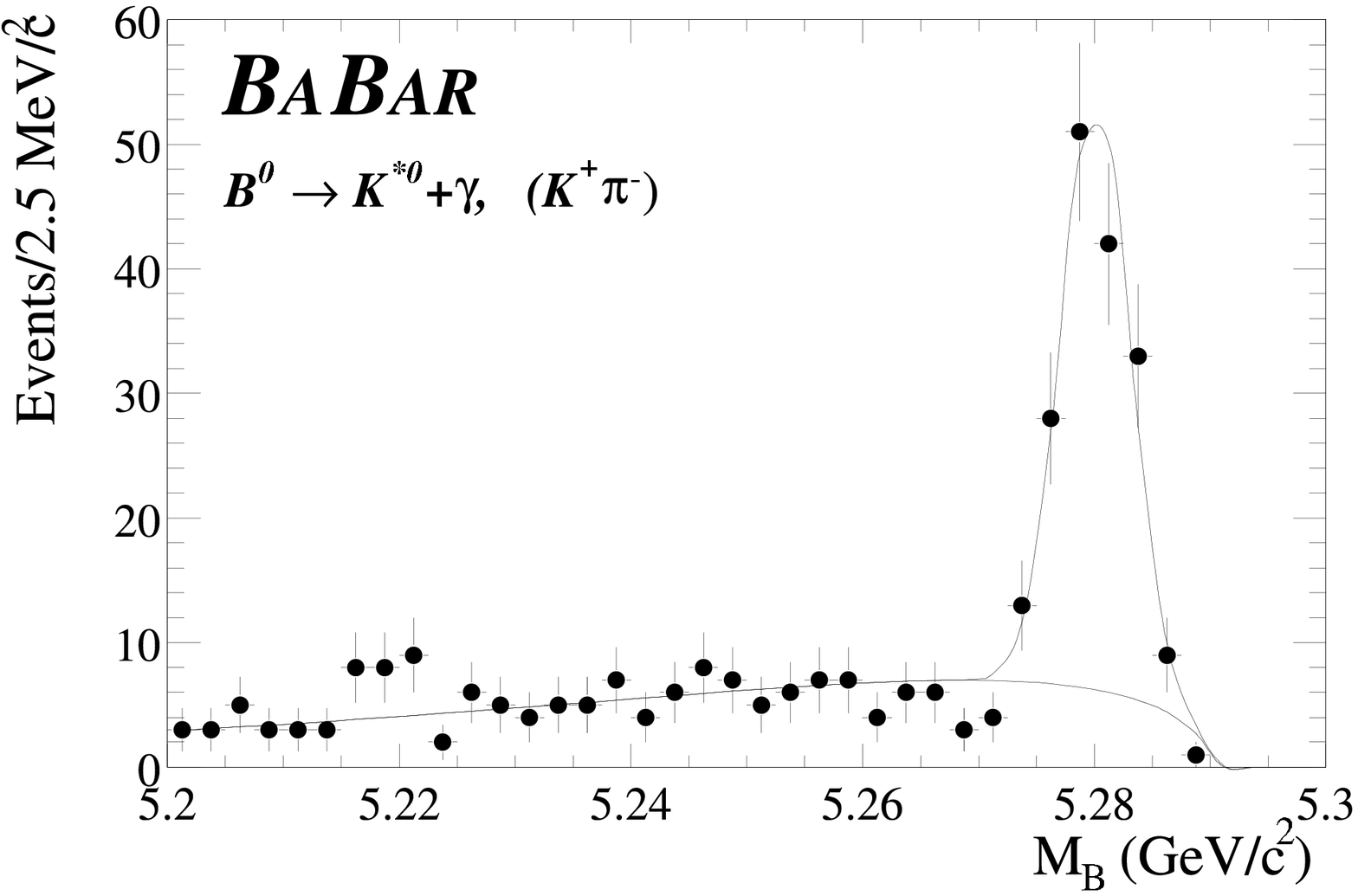}}
\kern-1.5in\lower 0.1in \hbox{(b)}
}
\end{center}
\caption{\it \label{fig:ksgmes}
The final sample of the \bksg\ search. 
a) is a scatter plot of $\Delta E$ vs. $m_{B}$. 
b) is a projection onto the $m_{B}$ axis, showing a clear 
\bksg signal.}
\end{figure}

\subsection{\bksg\ Results}
The efficiency for observing a \bksg\ event is calculated in Monte Carlo.
This includes corrections for known differences between data and Monte Carlo
for tracking efficiency, photon detection efficiency and particle 
identification efficiency. 
The efficiency is found to be $\epsilon = 0.209 \pm 0.013$, where the error
is purely systematic.
One can then use this efficiency to calculate the branching ratio as:
\begin{equation}
BR(\bksg) = {N_{sig} \over N_{B\bar{B}} \times \epsilon \times B_{K^*}},
\end{equation}
where $N_{B\bar{B}}$ is the number of $B\bar{B}$ pairs in the sample, 
and $B_{K^*}$ is the branching ratio for $K^{*0} \rightarrow K^-\pi^+$.
The systematic error comes mostly from data-derived efficiency corrections.
The preliminary result for the branching ratio is then:
\begin{equation}
BR(\bksg) = (4.39 \pm 0.41_{stat} \pm 0.27_{syst})\times 10^{-5}
\end{equation}

Measurement of a possible CP asymmetry in this decays is also of interest.
The asymmetry is defined as:
\begin{equation}
A_{CP} = {N(\bar{B^0}\!\rightarrow \bar{K^{*0}}\gamma) - 
	N(B^0\!\rightarrow K^{*0}\gamma) \over
          N(\bar{B^0}\!\rightarrow \bar{K^{*0}}\gamma) +
	N(B^0\!\rightarrow K^{*0}\gamma})
\end{equation}
Many of the systematic errors of the branching ratio measurement cancel out 
in this measurement and we are left with a preliminary measurement of:

\begin{equation}
A_{CP} = -0.035 \pm 0.094 \pm 0.022
\end{equation}

\subsection{Other \bksg\ Modes}
Figure \ref{fig:ksgother} shows
evidence for three other modes of $B\!\rightarrow K^{*}\gamma$:
$B^0\!\rightarrow K^{*0}\gamma,  K^{*0}\!\rightarrow \pi^0 K_s^0$;
$B^+\!\rightarrow K^{*+}\gamma,  K^{*+}\!\rightarrow K^+ \pi^0$;
$B^+\!\rightarrow K^{*+}\gamma,  K^{*+}\!\rightarrow K_s^0 \pi^+$.
Although clear signals are evident in each of these modes, their
efficiencies and systematic errors have not yet been fully evaluated.
For this reason, no branching ratios  or asymmetries will be presented for 
them at this time.

\begin{figure}
\begin{center}
\hbox{
\hbox{\epsfysize 2.3 truein \epsfbox{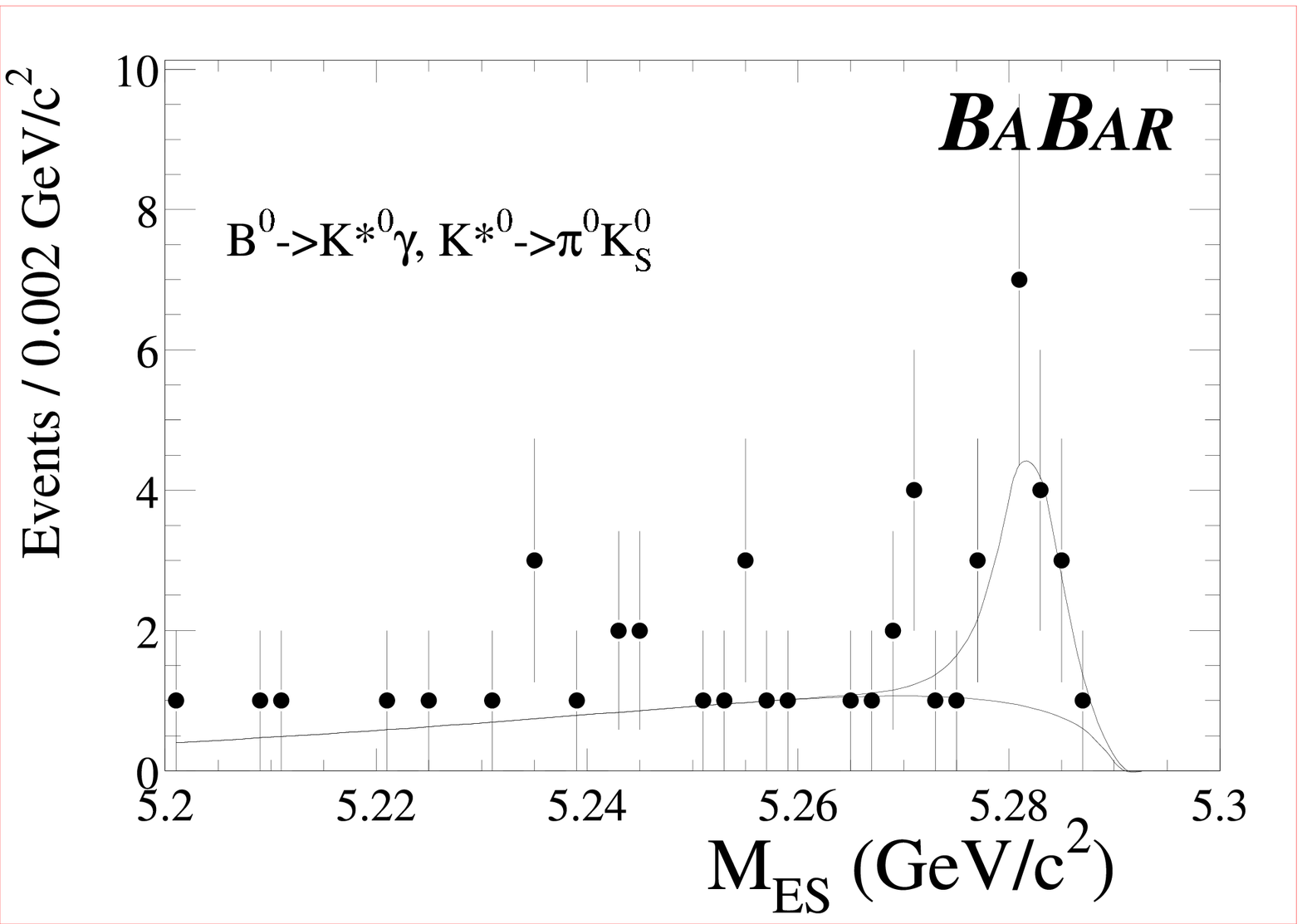}}
\hbox{\epsfysize 2.3 truein \epsfbox{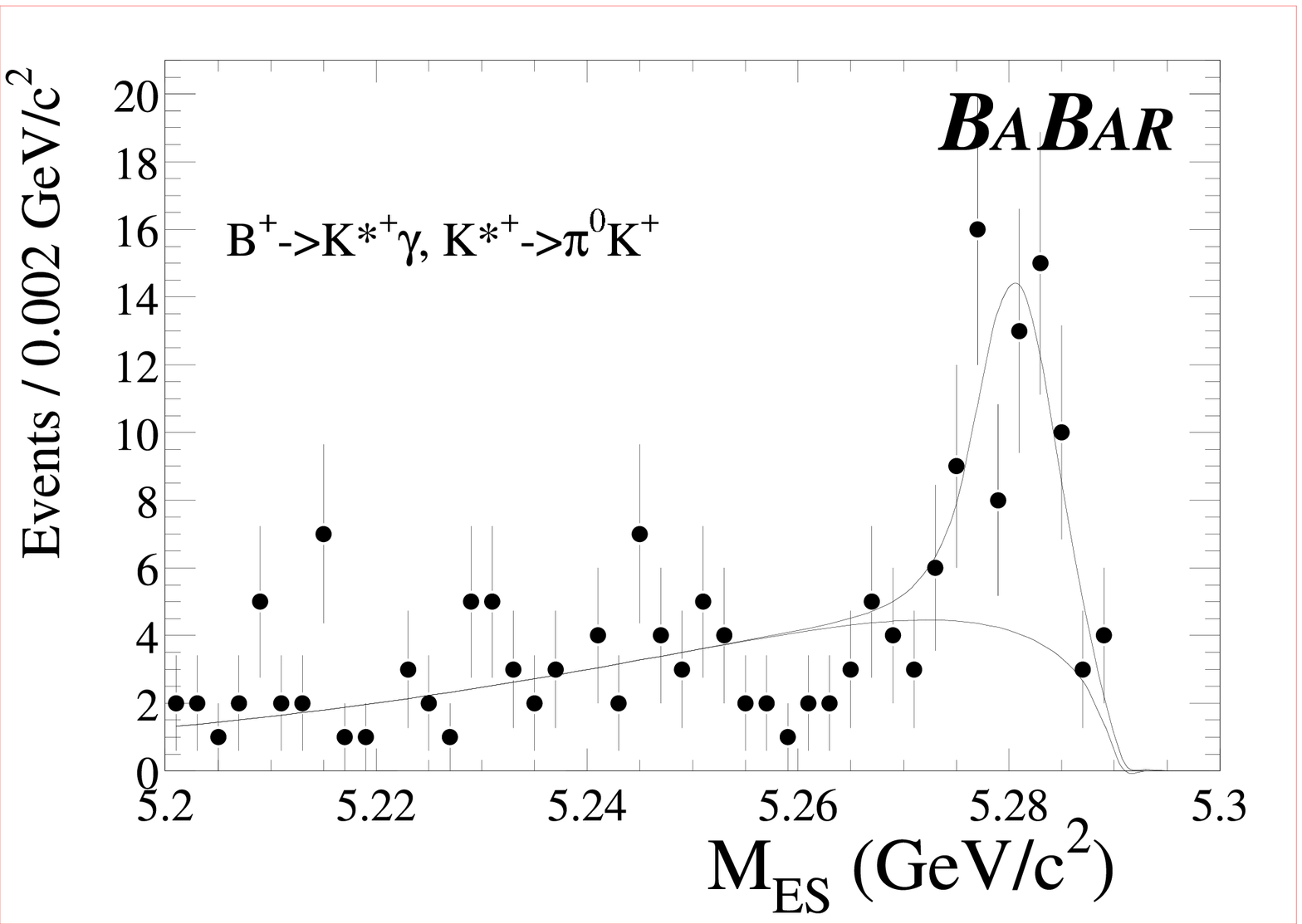}}
}
\end{center}
\begin{center}
\epsfysize 2.3 truein \epsfbox{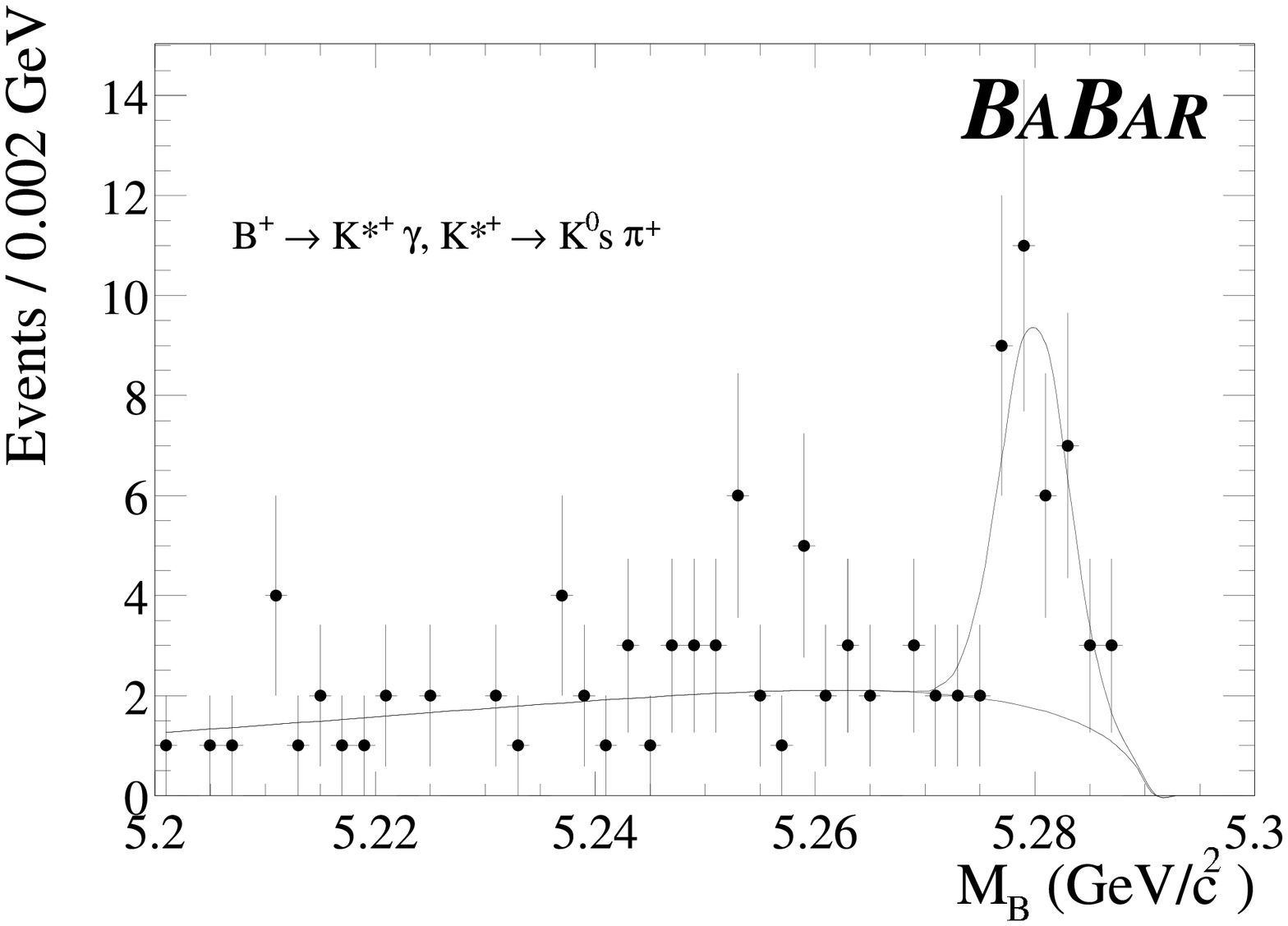}
\end{center}
\caption{\it \label{fig:ksgother} Signals for three other modes of
$B\!\rightarrow K^* \gamma$.}
\end{figure}

\section{Analysis of \bhh}
\subsection{\bhh Candidates}
\bhh candidates are composed out of pairs of oppositely signed
charged tracks, both of which must pass tracking quality cuts
and DIRC quality cuts.
The candidates must pass a cut on the ``sphericity angle''
which is almost identical to the ``thrust angle'', $\theta_t$ defined
in Section \ref{sec:bkgsupp},
of $\cos \theta_s < 0.9$.
Furthermore, the candidate events must pass cuts on the second
Fox-Wolfram moment of $R_2 < 0.95$ and on sphericity of
$S >0.01$. 
Finally, loose cuts are placed on the kinematic variables of the decay:
$5.2\, {\rm GeV} < m_{ES} < 5.3\, {\rm GeV}$ 
and $-0.15 \, {\rm GeV} < \Delta E < 0.15 \, {\rm GeV}$.
In Monte Carlo, these cuts are found to have an efficiency of
$\epsilon \approx 0.45$ for the \bpp mode.
The 26404 events that pass these cuts in the data sample are put into
a Maximum Likelihood fit described in the next section.

\subsection{\bhh Likelihood Fit}
Since a small number of signal events are expected in the \bhh modes, 
it is desirable to use a Maximum Likelihood fitting technique, rather
than the more traditional ``cut-and-count'' technique.
For the purposes of the fit, each event is considered to come from
one of eight hypotheses:
a true \bpp\ event, 
a true \bkp\ event (or charge conjugate), 
a true \bkk\ event, 
a background \bpp\ event, 
a background \bkp\ event (or charge conjugate), 
or a background  \bkk\ event.
The parameters that are varied in the
the likelihood fit are the number of events in each of these hypotheses.
The event variables that are fit to are listed in Table \ref{tab:likevar}.

\begin{table}
\begin{center}
\begin{tabular}{c|c}
Variable & Definition \\
\hline
$m_{ES}$   & Energy substituted B Mass, described in 
Section \ref{sec:kinvar} \\
$\Delta E$ &  Energy difference, described in 
Section \ref{sec:kinvar} \\
$F$  & Fisher discriminant output, described in Section
\ref{sec:bkgsupp} \\
$\theta_{C,+}$ & DIRC Cherenkov Angle for positive track \\
$\theta_{C,-}$ & DIRC Cherenkov Angle for negative track \\
\end{tabular}
\end{center}
\caption{\it \label{tab:likevar} Variables used in the Maximum Likelihood 
fit of \bhh\ and their definitions.}
\end{table}

The likelihood function is defined as follows.
A probability distribution function (PDF) is calculated for each event
and hypothesis.
It has the form:
\begin{equation}
P^{hypo}_{event} = P^{hypo}_{M_{ES}} 
	           P^{hypo}_{\Delta E}
                   P^{hypo}_{F}
                   P^{hypo}_{\theta_{C,+}}
                   P^{hypo}_{\theta_{C,-}},
\end{equation}
where the $P$'s are the PDF's for each variable given an event hypothesis.
Typically, these are parameterized in terms of simple functions.
The full likelihood function is then written,

\begin{equation}
{\cal L} = exp(-\sum_{hypo} N_{hypo})\prod_{j=1}^N
\big[\sum_{hypo}N_{hypo} P^{hypo}_{event}\big],
\end{equation}
where $N_{hypo}$ is the number of events in each hypothesis and 
$N$ is the total nmber of events in the sample.
This function is then maximized using standard tools.

\subsection{Calibration of PDF's}
In order for the the fit to be valid, it is essential that the 
individual PDF's properly describe the distributions for the
signal and background hypotheses.
These PDF's are calculated based on calibration samples of real 
data and checked with Monte Carlo simulation.
The following sections describe this procedure in more detail
for each of the five fit variables.

\subsubsection{$m_{ES}$ PDF's}
The shape of the $m_{ES}$ distribution for signal 
is taken to be Gaussian. 
Since its width is dominated by the beam spread energy, it can be 
reliably calculated based on fully reconstructed decays of the
type $B^- \!\rightarrow D^0(\!\rightarrow K^-\pi^+)\pi^-$.
Figure \ref{fig:mescal}(a) shows the shape of this distribution.
For background, the shape is assumed to be the ``Argus Function'',
which is fitted to data taken on the $\Upsilon(4s)$ peak, but with
$\Delta E$ outside of the signal region.
Figure \ref{fig:mescal}(b) shows this distribution and its fit.

\begin{figure}
\begin{center}
\hbox{
\hbox{\epsfysize 2.6 truein \epsfbox{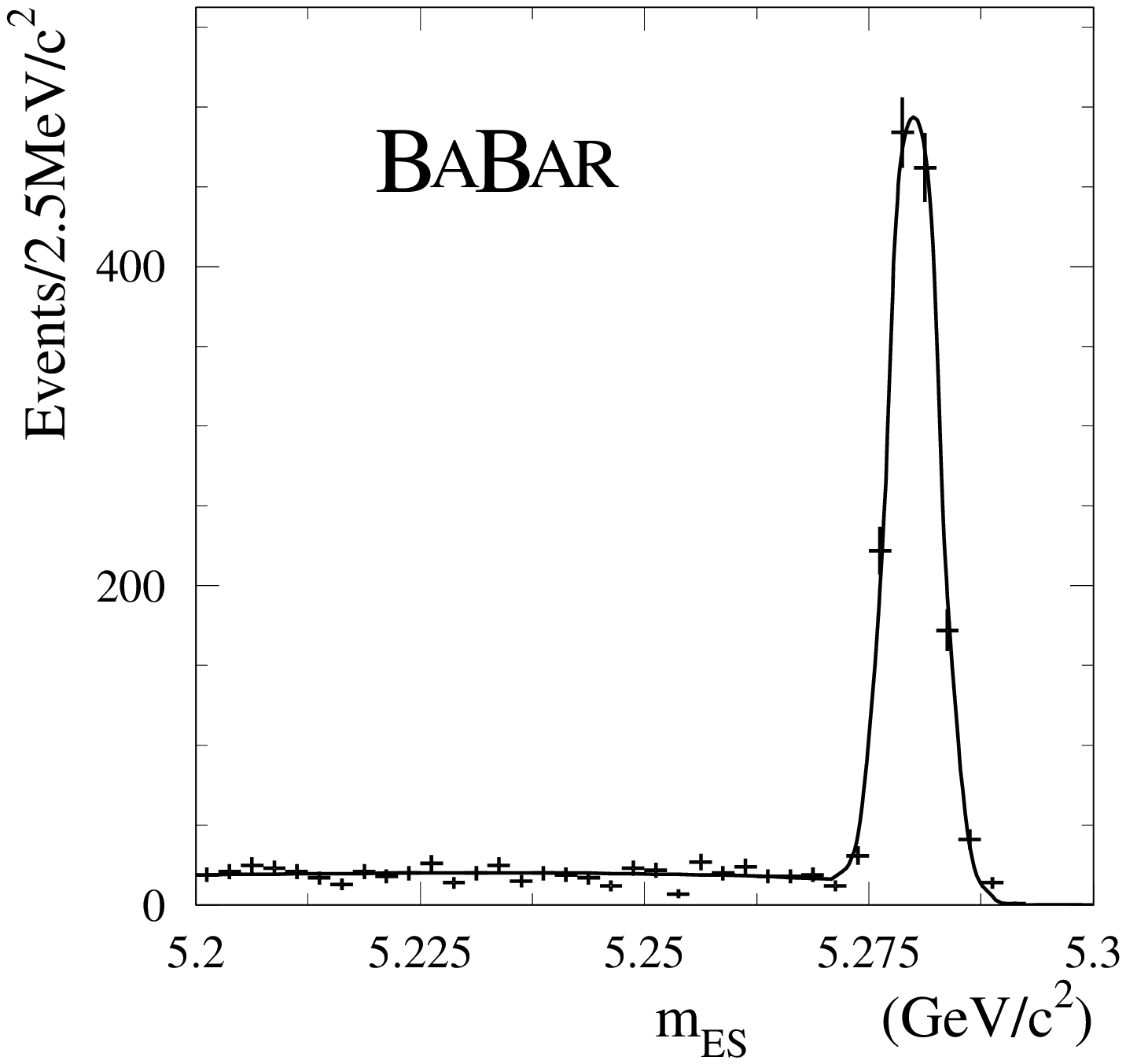}}
\kern-1.2in\lower 0.1in \hbox{(a)}
\kern1.8in
\hbox{\epsfysize 2.6 truein \epsfbox{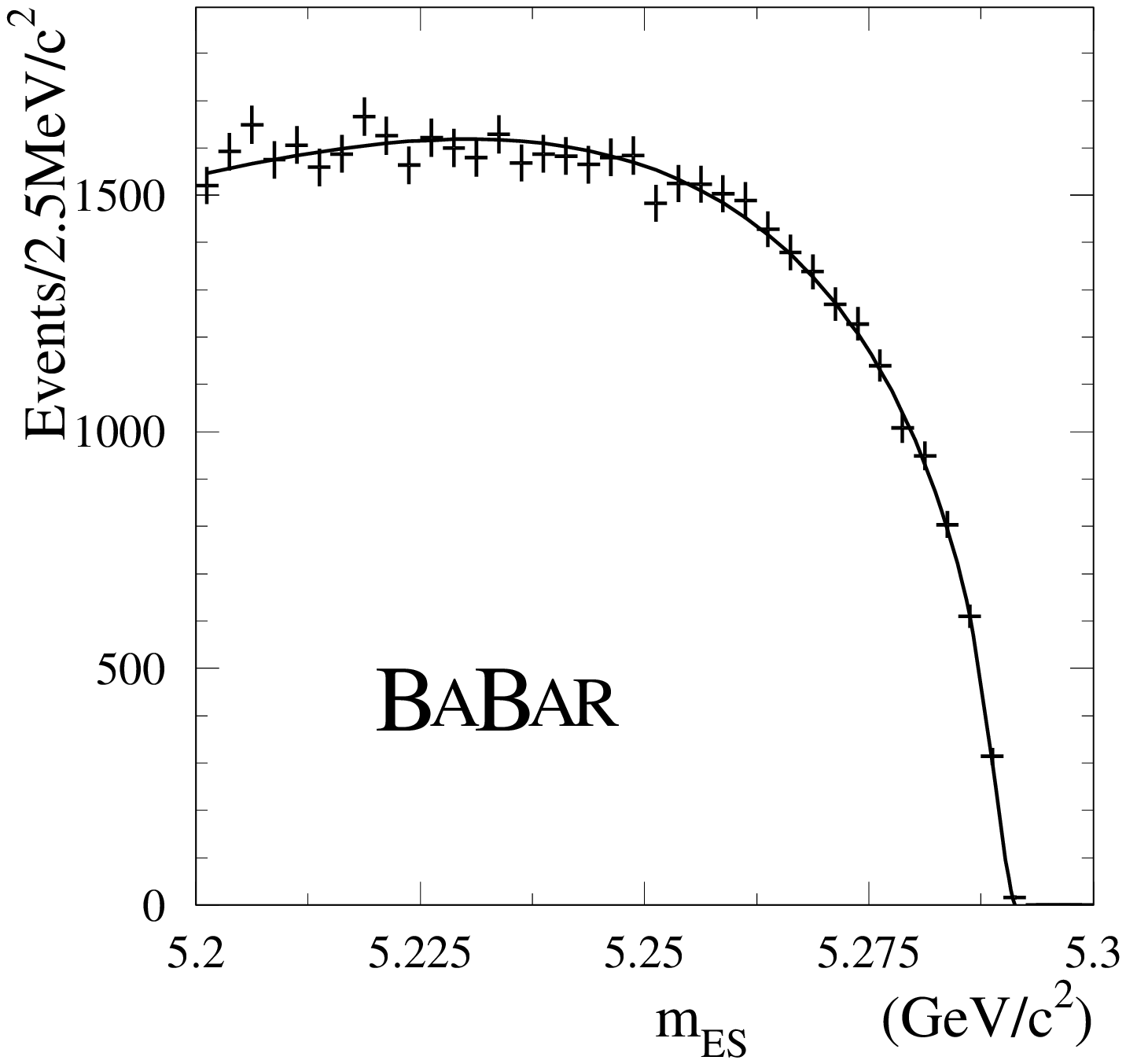}}
\kern-1.2in\lower 0.1in \hbox{(b)}
}
\end{center}
\caption{\it \label{fig:mescal}
$m_{ES}$ distributions used to calibrate the $m_{ES}$ PDF's.
a) is from $B^- \!\rightarrow D^0(\!\rightarrow K^-\pi^+)\pi^-$  data and
is used for the signal PDF.
b) is from $\Delta E $ sideband data and is used for the background PDF.
}
\end{figure}

\subsubsection{$\Delta E$ PDF's}
The $\Delta E$ shape for signal is assumed to be Gaussian. 
In contrast to the $m_{ES}$ case, however, the width is dominated by
tracking resolution and is therefore different for 
$B^- \!\rightarrow D^0(\!\rightarrow K^-\pi^+)\pi^-$ than for
\bhh.
In this case, we compare the width measured in data for 
$B^- \!\rightarrow D^0(\!\rightarrow K^-\pi^+)\pi^-$, 
$\sigma_{\Delta E} = 19 MeV$, with that found in Monte Carlo,
$\sigma_{\Delta E} = 15 MeV$.
We then scale up the width found in signal Monte Carlo 
($\sigma_{\Delta E} = 21 MeV$) by the same factor to obtain
the width used for the PDF, $\sigma_{\Delta E} = 26 \pm 5 MeV$.
For background, the shape is taken to be a polynomial, which is
fitted to data taken on peak, but outside of the $\Delta E$ signal
region.

\subsubsection{Fisher Discriminant PDF's}
Double Gaussians are used for the Fisher discriminant shape. 
Signal Monte Carlo is used to determine the parameters for signal.
It is checked with the
\hbox{$B^- \!\rightarrow D^0(\!\rightarrow K^-\pi^+)\pi^-$} sample.
For background, the parameters are determined using the $m_{ES}$ 
sideband and are checked using off-resonance data and continuum
Monte Carlo.

\subsubsection{$\theta_C$ PDF's}
A sample of 
$D^{*+} \!\rightarrow \pi_{slow}^+ D^0(D^0\!\rightarrow K^-\pi^+)$
was used to provide clean samples of $\pi^+$'s and $K^+$'s.
The DIRC $\theta_C$ response was parameterized by a Gaussian with mean 
and width dependent on dip angle.
Also, it was necessary to include small non-Gaussian ``satellite'' peaks
in order to adequately model the performance.

\subsection{\bhh\ Fit Result}
The maximum likelihood fit yields the results shown in Table 
\ref{tab:bhhres}.
The systematic errors are calculated by varying the parameters of the
PDF's, both within their statistical errors and to cover any 
disagreements between data and Monte Carlo.
More information about this analysis may be found in
\cite{twobodypp}.

\begin{table}
\begin{center}
\begin{tabular}{c|c|c}
Decay Mode & $N_{signal} \pm \sigma_{stat} \pm \sigma_{syst}$ &
BR BABAR ($\times 10^{-6}$) \\
\hline
$\pi^+\pi^-$ & $41 \pm 10 \pm 7$ & $4.1 \pm 1.0 \pm 0.7$ \\
$K^+\pi^-$ & $169 \pm 17 ^{+12} _{-17}$ & $16.7 \pm 1.6 ^{1.2} _{-1.7}$ \\
$K^+K^-$   & $8.2 ^{+7.8} _{-6.4} \pm 3.3$ &$ < 2.5 $(90\% C.L.) \\
\end{tabular}
\end{center}
\caption{\it \label{tab:bhhres} Number of events for each mode found 
in the Maximum Likelihood Fit and the corresponding measured
branching ratios.}
\end{table}

\section{Conclusions}
Based on a first year sample of $22.4\times 10^6$ $B\bar{B}$ pairs, 
BABAR has the preliminary measurements shown in Table \ref{tab:concl}.
As more data is collected and more decays modes analyzed, BABAR will
have many more results on charmless B decays.

\begin{table}
\begin{center}
\begin{tabular}{c|c}
Parameter & Measurement \\
\hline
BR(\bksg) & $(4.39 \pm 0.41 \pm 0.27) \times 10^{-5}$\\
$A_{CP}$(\bksg) & $-0.035 \pm 0.094 \pm 0.022$ \\
BR(\bkp) &  $(16.7 \pm 1.6 ^{+1.2} _{-1.7}) \times 10^{-6}$ \\
BR(\bpp) &  $(4.1 \pm 1.0 \pm 0.7) \times 10^{-6}$ \\
\end{tabular}
\end{center}
\caption{\it \label{tab:concl} Summary of measurement presented in this
paper. In each case, the first error is statistical and the second s
systematic.}
\end{table}

\end{document}